%% file: MAIN.tex
\documentclass[letterpaper,twocolumn,10pt]{article}
\usepackage{usenix2020}

\usepackage{tikz}
\usepackage{amsmath,shortcuts}

\begin{document}
\date{}

\pagenumbering{gobble}

\title{\Large \bf \Title}

\ifdefined\showAuthors
\author{
{\rm Avinash Sudhodanan}\thanks{Supported by a research grant from the Microsoft Security Response Center (MSRC).}\\
Independent Researcher
\and
{\rm Andrew Paverd}\\
Microsoft Security Response Center
} 
\fi

\maketitle

\input{abstract}
\input{intro}
\input{background}
\input{threat}
\input{attacks}

\input{experiments}
\input{results}
\input{case-studies}

\input{discussion}

\input{related}
\input{conclusion}
\raggedbottom

\ifdefined\showAuthors
\input{ack}
\fi


\raggedright
\bibliographystyle{plain}
\bibliography{../related-work/references}

\end{document}

%% file: abstract.tex
\begin{abstract}

The ubiquity of user accounts in websites and online services makes \emph{account hijacking} a serious security concern.
Although previous research has studied various techniques through which an attacker can gain access to a victim's account, relatively little attention has been directed towards the process of \emph{account creation}.
The current trend towards federated authentication (e.g., Single Sign-On) adds an additional layer of complexity because many services now support both the classic approach in which the user directly sets a password, and the federated approach in which the user authenticates via an identity provider.

Inspired by previous work on preemptive account hijacking~\cite{GhasemisharifSSignOff2018USENIX}, we show that there exists a whole class of \emph{account pre-hijacking} attacks.
The distinctive feature of these attacks is that the attacker performs some action \emph{before} the victim creates an account, which makes it trivial for the attacker to gain access \emph{after} the victim has created/recovered the account.
Assuming a realistic attacker who knows only the victim's email address, we identify and discuss \nNovelAtkTypes{} different types of account pre-hijacking attacks.

To ascertain the prevalence of such vulnerabilities in the wild, we analyzed \nSitesTested{} popular services and found that at least \nSitesVuln{} of these were vulnerable to one or more account pre-hijacking attacks.
Whilst some of these may be noticed by attentive users, others were completely undetectable from the victim's perspective.
Finally, we investigated the root cause of these vulnerabilities and present a set of security requirements to prevent such vulnerabilities arising in future.

\end{abstract}

%% file: intro.tex
\section{Introduction}
\label{sec:intro}

User accounts have become a ubiquitous feature of websites and other online services.
Correspondingly, such accounts have become valuable targets for attackers, and companies invest significant resources to prevent \emph{account hijacking} attacks, in which an attacker gains unauthorized access to the victim's account.
Previous work on this topic has studied various techniques that could be used for account hijacking, for example, the use of Cross-Site Request Forgery (CSRF) to trick victims into changing their account passwords to an attacker-controlled value~\cite{PortSwiggerCSRF, CSRF, SudhodananAuthCSRF2017EuroSnP}.

In an effort to improve user experience, there is currently a trend towards federated identity and authentication.
One of the most visible aspects of this is Single Sign-On (SSO) in which the user creates an account with an Identity Provider (IdP), and can then use this to create accounts with any relying party (RP) service that supports SSO and trusts the user's IdP.
There is a strong incentive for services to support SSO because it improves the experience for users by allowing them to reuse the same IdP account across multiple services.
Many large companies, including Google, Facebook, and Microsoft, provide IdP services that are widely supported and trusted by websites and other online services.
Previous work has also explored the security implications of SSO, showing that IdPs can become single points of failure~\cite{JainAppleSSOBug, BaikarFacebookSSOBug,GhasemisharifSSignOff2018USENIX,WangWebSSOBrm2012IEEESP,BaiAuthScan2013NDSS,SunOAuthSec2012CCS}.

However, one aspect that has not received much attention is the process of \emph{account creation}, along with its corresponding security assumptions and requirements.
This process is further complicated by the move towards SSO because many services now support two different mechanisms through which users can create an account: the \emph{classic} approach in which the user sets a password directly with the service, and the \emph{federated} approach where the user already has an account with an IdP and uses this to create an account with the service.
Once an account has been created, some services also offer the possibility to \emph{link} an IdP account, so that the user can either sign in directly to the service or authenticate via the IdP.

Ghasemisharif et al.~\cite{GhasemisharifSSignOff2018USENIX} presented the first example of a \emph{preemptive account hijacking} attack, in which an attacker gains control of a victim's federated identity (e.g., the victim's IdP account) and uses this to create accounts at services for which the victim has not yet signed up.
The attacker then waits for the victim to join that service and start using their ``new'' account.
At a later time, the attacker can sign into the service using the compromised IdP account and view or manipulate any information stored by the victim in that account.

Inspired by that attack, we demonstrate that there exists an entire class of such attacks, which we call \emph{account pre-hijacking} attacks.
The distinguishing feature of these attacks is that the attacker performs some action \emph{before} the victim creates an account at the target service.
The unsuspecting victim might subsequently create/restore their account and start using it.
Finally, the attacker completes the attack by gaining access to the victim's account.
In this work, we identify and describe \nNovelAtkTypes{} types of pre-hijacking attacks\footnote{However, we do not claim that this is an exhaustive list of attacks.}.
In contrast to the attack by Ghasemisharif et al.~\cite{GhasemisharifSSignOff2018USENIX}, none of our attacks require the attacker to compromise the victim's IdP account.

All of our attacks make some assumptions about the victim's actions or lack thereof.
These include (common) actions, such as creating an account (using the classic or federated route) or recovering the password for an existing account, as well as inactions, such as ignoring emails from services where the user does not have an account, or failing to notice unexpected identifiers after recovering an account. 
One variant requires a successful CSRF attack.
We set out the assumptions for each type of attack in Table~\ref{tab:assumptions}.

An interesting aspect of account pre-hijacking attacks is that they require the attacker to anticipate which services the victim is likely to sign up for and take action \emph{before} the victim creates an account.
This could be achieved in various different ways.
For example, the attacker may already know which services a specific victim uses, and opportunistically pre-hijack accounts at other services the victim is likely to use.
More broadly, the attacker might learn that a whole organization (e.g., a university department) plans to use a particular service, and pre-hijack accounts for any publicly-known email addresses from that organization.
Alternatively, the attacker may observe a general increase in popularity of a service (e.g., a video conferencing service when people are forced to work from home) and pre-hijack accounts for any email addresses found through e.g., website scraping.
There is no risk to the attacker if the victim has already created an account. 

To ascertain the prevalence of services that are vulnerable to account pre-hijacking attacks, we analyzed \nSitesTested{} of the most popular websites, based on the Alexa global website rankings~\cite{alexatop500}.
We found that at least \nSitesVuln{} of these were vulnerable to one or more pre-hijacking attacks, including widely-used cloud storage, social and professional networking, blogging, and video conferencing services.
We responsibly disclosed these vulnerabilities to the affected organizations and our reports have already been acknowledged as being of \emph{high} severity by a major video conferencing service.

Fundamentally, the root cause of these attacks is that the service or IdP fails to verify that the user actually owns the supplied identifier before allowing use of the account.
Although many services do perform this verification (e.g., sending a confirmation code or URL to the provided email address), the vast majority allow the user to begin using the account \emph{before} this verification has been completed, thus opening a window of vulnerability for account pre-hijacking attacks.
We hope that our work will serve to underline the importance of such verification by highlighting the consequences of its omission.
However, recognizing that this may not be immediately practicable for all services, we also present a set of defense in depth security requirements for account creation that would have mitigated all the vulnerabilities we identified.

In summary, we make the following contributions:
\begin{itemize}
	\item We investigate security failures in user account creation and identify a novel class of \emph{account pre-hijacking} attacks in which the attacker takes steps to compromise a user account \emph{before} the user has created the account. We describe \nNovelAtkTypes{} specific types of account pre-hijacking that are applicable to current websites and online services.
	\item We analyzed \nSitesTested{} popular services to ascertain the prevalence of vulnerabilities that could be exploited in account pre-hijacking attacks. We found that at least \nSitesVuln{} services were vulnerable to one or more attacks. We  disclosed these vulnerabilities to the respective organizations and our reports have already been acknowledged as being of high severity by a major video conferencing service.
	\item We found that the root cause of these attacks is failure to verify ownership of the identifier \emph{before} allowing the account to be used. We also present an additional set of defense in depth security requirements for account creation and management that would have mitigated all the vulnerabilities we identified.
\end{itemize}

%% file: background.tex
\section{Background}
\label{sec:back}

\subsection{Single Sign-On (SSO)}
\label{sec:sso}

SSO is one of the most prevalent examples of federated identity management.
In SSO, the user maintains a single account with an Identity Provider (IdP) and uses this to create accounts and authenticate to one or more Relying Party (RP) services.

When the user visits the RP and initiates the authentication flow, the RP redirects the user to the IdP's SSO-authentication end-point.
The user then authenticates to the IdP, using their IdP account, and gives consent for the IdP to send specific details to the RP.
If the authentication is successful, the IdP sends a proof-of-authentication to the RP, such as the authentication assertion in SAML 2.0~\cite{saml2}, or the authorization code (authentication token) in OAuth 2.0~\cite{OAuth}.
The RP can use this proof to fetch additional attributes about the user from the IdP, such as the user's email address.
If the user does not already have an account with the RP service (i.e., a new user), the RP service uses the attributes provided by the IdP to create an account.
The service may optionally request additional information from the user during account creation.
Importantly, the RP records which IdP is associated with the account, and will expect a proof-of-authentication from the same IdP when the user attempts to authenticate in the future.

Federated identity is becoming increasingly popular as it reduces the burden on users of having to create, remember, and manage separate authentication credentials for each service.
It also offers security benefits as IdPs tend to invest in securing users' accounts, e.g., using multi-factor authentication.

\subsection{Authentication Attributes}
\label{sec:authattr}

\begin{figure}
	\centering \includegraphics[scale=.5]{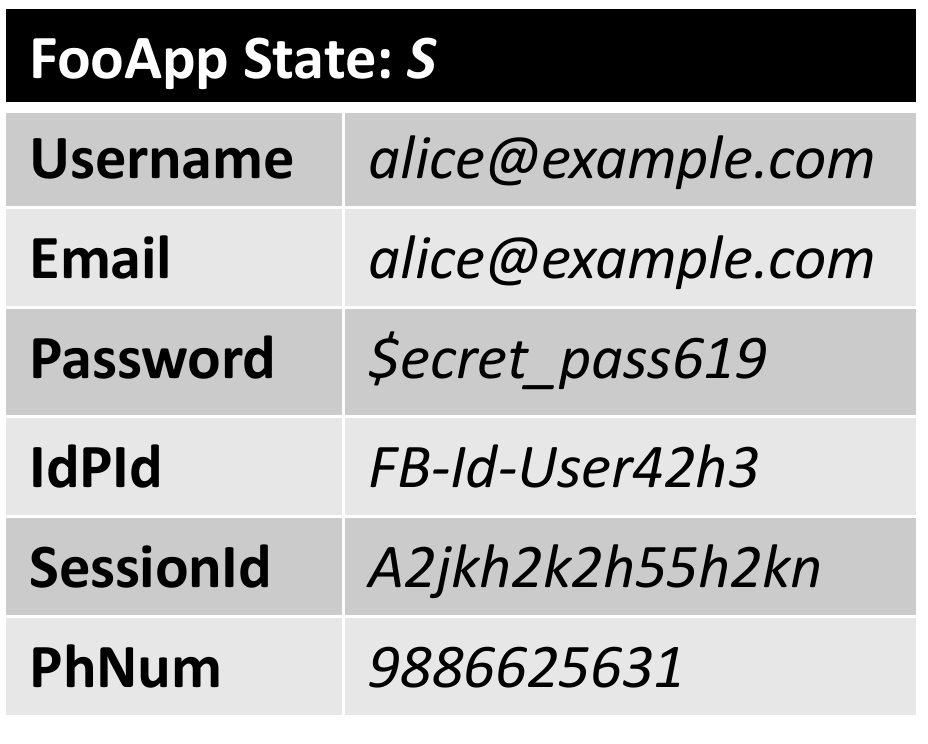}
	\caption{Example state of the authentication attributes for an account at the FooApp service.}
	\label{fig:ssas}
\end{figure}

User accounts necessarily include various pieces of information for authenticating the user -- we refer to these as \emph{authentication attributes}.
Some of these attributes may be persistent, such as the user's username and password, whilst others may be transient, such as the list of currently valid authentication cookies.
For a given account, we refer to the set of authentication attributes and their corresponding values as the \emph{state} of the account at that specific point in time.
Figure~\ref{fig:ssas} shows an example of such a state.

In this example, the \textsf{Username} attribute of this account is the user's email address.
This need not always be the case, but a large number of online services use email addresses as usernames as these are guaranteed to be unique, and are easy for users to remember.
In this paper, we assume that the user's email address is always used as the account username.

The \textsf{Email} and \textsf{PhNum} (phone number) attributes can also be used to recover access to the account if the user forgets the password.
Although implementations may differ, the general pattern for a password reset is for the service to send a secret capability (e.g., a code or a URL with an embedded authentication token) to the registered email address or phone number.
Using this capability, the legitimate user can authenticate to the service and reset the password.

As expected, the \textsf{Password} attribute represents the user-chosen secret used to authenticate the user to the service.
Although Figure~\ref{fig:ssas} shows a representative example, real services should store the password securely (e.g., storing only a salted hash of the password).
Although not shown, this attribute also encompasses any other authentication secrets, such as secret keys used in multi-factor authentication (MFA).

The \textsf{IdPId} attribute records the identity of the user as provided by a federated identity provider.
Depending on the service, it may be possible to add one or more federated identities to an existing account, after the account has been created (e.g., the option to \emph{``Connect with Facebook''} or an equivalent IdP).
In our representation, this would be recorded as a change to the value of the IdP attribute.

The \textsf{SessionId} attribute encompasses all currently-active sessions' identifiers (e.g., valid authentication cookies) for the account.
This is a transient attribute as sessions usually end when the user signs out, or after a period of inactivity.

%% file: threat.tex
\section{Threat Model}
\label{sec:threat}

The attacker's goal is to gain control (i.e., \emph{hijack}) the victim's user account at the target service.
Depending on the nature of the service, this could allow the attacker to access the victim's confidential information (e.g., messages, documents, billing statements, etc.) or to impersonate the victim (e.g., sending messages, subscribing to services, etc.).

We assume the same \emph{web attacker} threat model used in prior work e.g.,~\cite{AkhaweFormal2010CSF, Mainka2016MalIdp, SudhodananMPWA2016NDSS}, in which the attacker can access the target service as well as third-party IdP services such as those provided by Google, Facebook, and Microsoft.
The attacker can create both free and paid user accounts at the target service, but does not have administrative rights at this service.
The attacker can also create accounts at one or more IdPs and use these with the target service.
If the target service supports custom IdP integration, the adversary can select any publicly available identity management service (e.g., OneLogin \cite{OneLogin}) and use this with the target service.

Additionally, we assume that the attacker knows the email address and other basic details of the victim (e.g., first and last name).
These details would already be known to the attacker in the case of a targeted attack, or might be e.g., scraped from social media or found in password database dumps in the case of an untargeted attack. 
The attacker can use these details to create accounts both at the target service and IdP.

For some attacks, we assume the attacker can make the victim visit an attacker-controlled URL (e.g., through click-baiting~\cite{CBait}).
The resulting HTTP request from the victim's web browser may include HTTP Cookies~\cite{BarthCookies} within the constraints of the SameSite policy~\cite{SameSite} of the web browser and the destination endpoint. 

Finally, we assume the victim to have at least a basic level of security awareness.
This means the attacker \emph{cannot} perform successful phishing attacks on the victim. However, we assume that the victim ignores notifications sent by services where they do not have an account. This is realistic as prior work (e.g., \cite{Akhawe2013WarningUSENIX}) has shown the ineffectiveness of notifications.
We assume that the victims and services regularly update their software, and implement mitigations against software attacks such as code injection~\cite{Injection} and memory corruption~\cite{MemoryCorrupt}.

%% file: attacks.tex
\section{Account Pre-Hijacking Attacks}
\label{sec:attacks}

As with account hijacking, the attacker's goal in account pre-hijacking is to gain access to the victim's account.
The attacker may also care about the \emph{stealthiness} of the attack, if the goal is to remain undetected by the victim.

The impact of account pre-hijacking attacks is the same as that of account hijacking.
Depending on the nature of the target service, a successful attack could allow the attacker to read/modify sensitive information associated with the account (e.g., messages, billing statements, usage history, etc.) or perform actions using the victim's identity (e.g., send spoofed messages, make purchases using saved payment methods, etc.).
We provide specific examples of the attack impact for the five case studies we describe in Section~\ref{sec:cases}.
Abstractly, an account pre-hijacking attack consists of three phases: 1)~\phaseOne, 2)~\phaseTwo, and 3)~\phaseThree.

\textbf{1. \phaseOne.} In the first phase, the attacker performs some preparatory action, such as creating an account at the target service using an identifier belonging to the victim (e.g., the victim’s email address, mobile phone number, etc.).
This requires the attacker to know the victim's identifier, but is not unrealistic since these identifiers are typically not secret.
As explained in Section~\ref{sec:threat}, victims' email addresses can be scraped in bulk for untargeted attacks, or would already be known by the attacker in the case of a targeted attack.
The main requirement in this phase is that the victim \emph{must not have created an account with the target service using that identifier}.
By definition, this would prevent account pre-hijacking.
It is easy for the attacker to detect if this is the case (e.g., the service will respond saying the account already exists), and this typically poses no risk to the attacker.
As discussed in Section~\ref{sec:intro}, the success of this attack depends on the attacker's ability to \emph{anticipate} the services at which the victims will create accounts.
At the end of this phase, the attacker waits for the victim to complete the second phase.

\textbf{2. \phaseTwo.} In the second phase, the victim either creates or recovers their account at the target service using the same identifier used by the attacker in the pre-hijacking phase.
Ideally, this would be the point at which the victim realizes that pre-hijacking has taken place, based on some type of notification from the service.
However, as discussed with the concrete attacks later in this section, the type of notification that can be shown to the victim depends on the specific type of pre-hijacking attack. 
Furthermore, as discussed in Section~\ref{sec:exp}, we observed that vulnerable services vary significantly in terms of the notifications they provide, ranging from providing warnings that \emph{might} tip off security-conscious users, to providing no notifications at all.
Assuming the victim is not notified or does not heed the notification, they would continue using their account as normal.

\textbf{3. \phaseThree.} In the third phase, the attacker gains access to the victim's pre-hijacked account.
The specific techniques used by the attacker in this phase depend on which technique was used in the first phase.
Some techniques used in this phase cause the victim to lose access to their account (i.e., low stealthiness), whereas others allow the victim and attacker to access the account concurrently (i.e., high stealthiness).

\begin{figure*}[ht!]
	\centering
	\includegraphics[width=\textwidth]{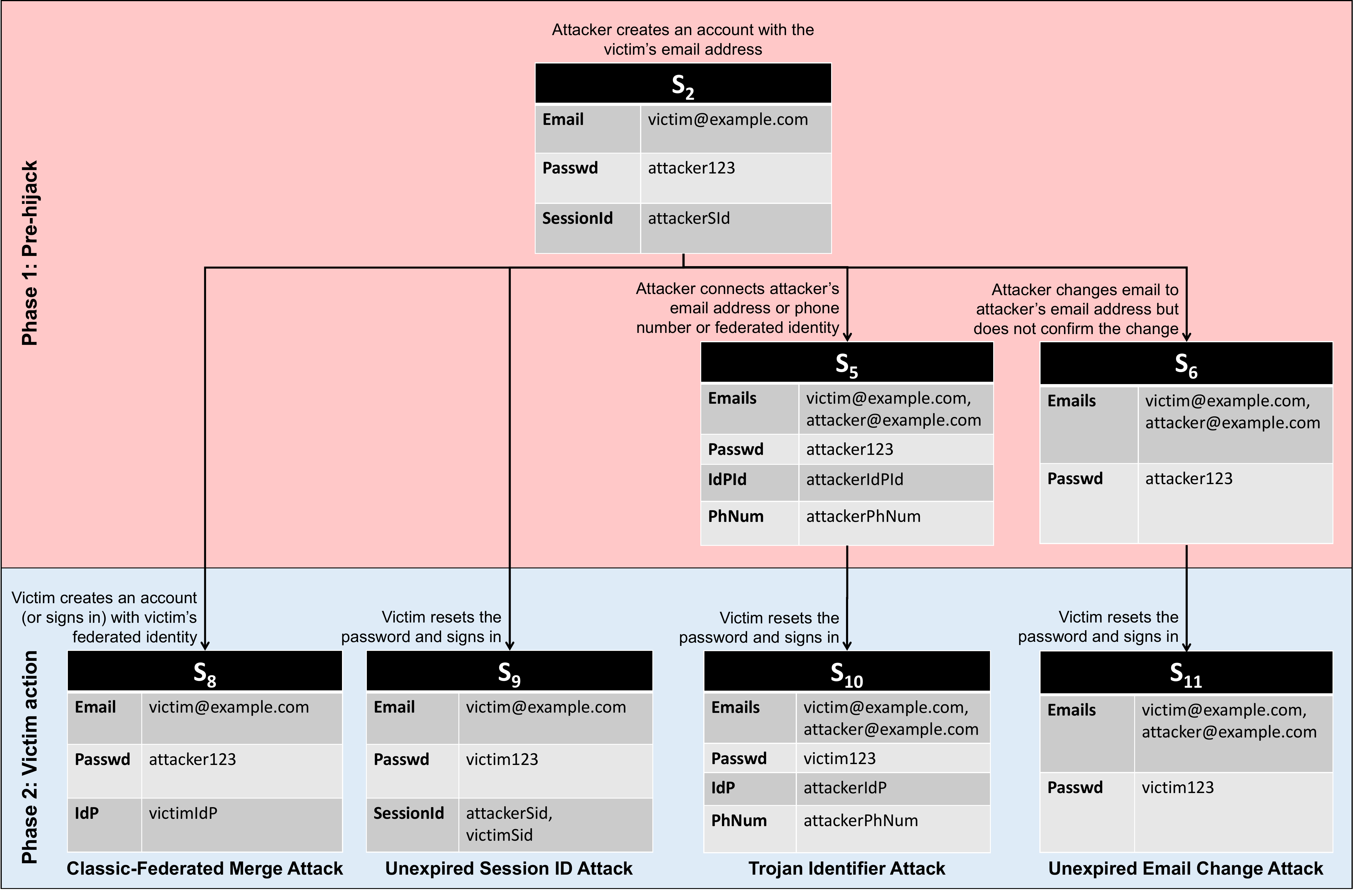}
	\caption{Attack tree showing the first two phases of the \classFedMerge (Section~\ref{sec:atk1}), \unexpSess (Section~\ref{sec:atk2}), \trojId (Section~\ref{sec:atk3}), and \unexpEmail (Section~\ref{sec:atk5}) attacks.}
	\label{fig:attacks1}
\end{figure*}

\begin{figure*}[ht!]
	\centering
	\includegraphics[width=\textwidth]{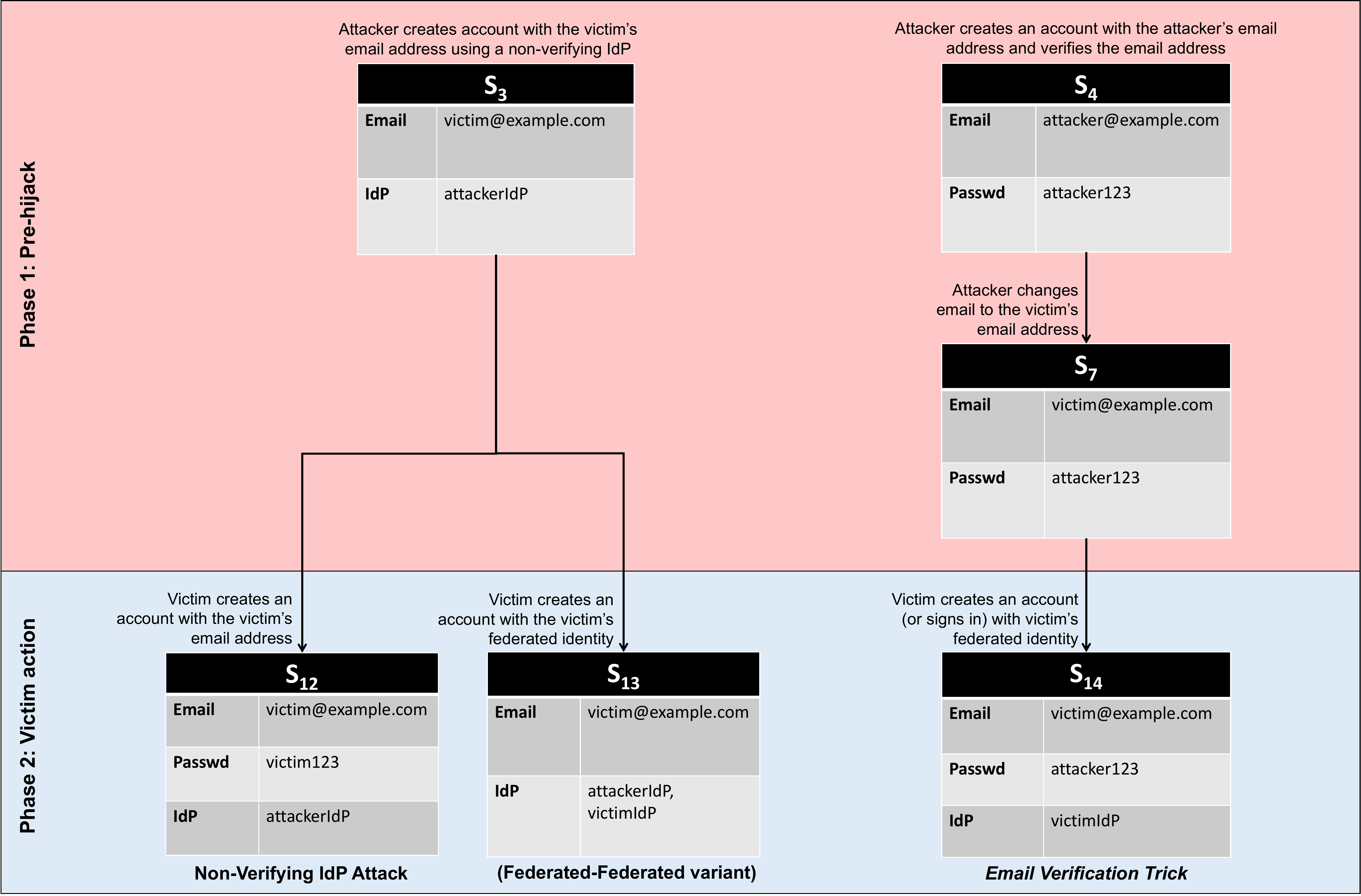}
	\caption{Attack tree showing the first two phases of the \nonverifIdp attack (Section~\ref{sec:atk5}) and the Email Verification trick (Section~\ref{sec:emailVerifVariation}) applied to the \classFedMerge Attack.}
	\label{fig:attacks2}
\end{figure*}

In the following subsections, we describe \nNovelAtkTypes{} concrete account pre-hijacking attacks.
We do not claim that this is an exhaustive list of such attacks, but rather that these are the types of attacks we found to be possible in real services (as described in Section~\ref{sec:exp}).
The attack trees in Figures~\ref{fig:attacks1} and \ref{fig:attacks2} summarize these concrete attacks, showing both the actions of the attacker and victim, as well as the state of the authentication attributes for the account (\textbf{S$_2$} - \textbf{S$_{14}$}) after each action.
In all cases, the starting state (\textbf{S$_1$}) is empty, since the account has not yet been created.

In the descriptions of the concrete attacks below, for clarity of explanation, we use the victim's email address as the identifier since this is the most widely-used identifier in real services.
However, the same attacks may also be possible using other identifiers, such as the victim's mobile number.
Table~\ref{tab:assumptions} summarizes the assumptions we make for each attack.

\input{tables/tbl_assumptions}

\input{attack1}
\input{attack2}
\input{attack3}

\input{attack4}

\input{attack5}
\input{attack7}

%% file: tables/tbl_assumptions.tex
\begin{table*}
	\footnotesize
	\centering
	\renewcommand{\arraystretch}{1.0}
	\begin{tabularx}{\textwidth}{l X c c c c c}
		\toprule
		\textbf{Entity} & \textbf{Assumptions} & \textbf{CFM} & \textbf{US} & \textbf{TID} & \textbf{UE} & \textbf{NV} \\
		\midrule
		\multirow{12}{*}{Target Service}	
		& - does not allow multiple accounts with the same identifier to exist concurrently & \yes & \yes & \yes & \yes & \yes \\
		& - supports password reset functionality & \no & \yes & \yes & \yes & \no \\
		& - supports both classic and federated sign in & \yes & \no & \yes & \no & \var \\
		& - uses email address as an identifier & \yes & \no & \no & \no & \yes \\
		& - merges classic and federated accounts with the same identifier & \yes & \no & \no & \no & \yes \\
		& - supports both classic and federated sign in and sign up & \yes & \no & \no & \no & \yes \\
		& - supports concurrent sessions & \no & \yes & \no & \no & \no \\
		& - does not expire past sessions by default upon password reset & \no & \yes & \no & \no & \no \\
		& - allows associating a federated account to a classic account & \no & \no & \yes & \no & \no \\
		& - supports email-change functionality & \no & \no & \no & \yes & \no \\
		& - sends email-change capability URLs and does not expire them upon password reset & \no & \no & \no & \yes & \no \\
		& - supports federated sign in through at least one non-verifying IdP & \no & \no & \no & \no & \yes \\
		\midrule
		
		\multirow{8}{*}{Victim}
		& - ignores the emails sent by the services where they do not have an account & \yes & \yes & \yes & \yes & \yes \\	
		& - will attempt to create an account at the target service through the classic route & \no & \yes & \yes & \yes & \var \\
		& - will recover the account that already exists at the target service with victim's identifier & \no & \yes & \yes & \yes & \no \\
		& - will not check whether an unknown identifier is associated to the recovered account & \no & \no & \yes & \no & \yes \\
		& - will attempt to create an account at the target service through the federated route & \yes & \no & \no & \no & \var \\
		& - will not invalidate all the active sessions of the recovered account & \no & \yes & \no & \no & \no \\
		& - will not check whether an email-change verification is pending at the recovered account & \no & \no & \no & \yes & \no \\
		& - is susceptible to CSRF attacks while logged in at the target service & \no & \no & \no & \var & \no \\ 
		\midrule
		
		\multirow{4}{*}{Attacker}
		& - can create an account before the victim creates their account at the target service & \yes & \yes & \yes & \yes & \yes \\	
		& - knows the identifier used by the victim for creating accounts & \yes & \yes & \yes & \yes & \yes \\
		& - knows the email address of the victim's federated account & \yes & \no & \no & \no & \no \\
		& - can create an account with any email address at a non-verifying IdP & \no & \no & \no & \no & \yes \\
		\midrule
	\end{tabularx}
	\caption{Assumptions on the target service, victim, and attacker for the various Account Pre-Hijacking Attacks. \yes indicates that the assumption applies to the attack, \var indicates that it applies only to some variants of the attack, and \no indicates that the assumption is not necessary for the attack. The attacks are \classFedMerge (CFM), \unexpSess (US), \trojId (TID), \unexpEmail (UE), and \nonverifIdp (NV).}
	\label{tab:assumptions}
\end{table*}

%% file: attack1.tex
\subsection{\classFedMerge Attack}
\label{sec:atk1}

This attack exploits a potential weakness in the interaction between the classic (i.e., setting a password) and federated (i.e., SSO) approaches for account creation.
Specifically, the attacker uses the victim's email address to create an account via the classic approach, and the victim subsequently creates an account via the federated approach.
If not carefully handled, this could result in both the victim and the attacker having access to the same account.

\textbf{Preconditions.} The target service must support both classic and federated account creation, and should use email addresses as the unique account identifiers (i.e., in place of usernames).
Furthermore, the service must allow users to associate a federated identity (e.g., SSO) with an existing (email and password) account -- a property we refer to as \emph{IdP account connect}.
Services typically achieve this by checking that the email address provided in the assertion from the IdP matches the email address used during classic account creation. 

\textbf{1. \phaseOne.} The attacker creates an account at the target service via the classic route, but provides the email address of the victim and a password of the attacker's choosing.
Specifically, the attacker provides the email address they anticipate the victim will use for federated authentication.
At this point, the target service might send a confirmation email to the victim's email address, asking the recipient to confirm their email address (e.g., either by clicking on a link or entering the code provided).
However, the victim might ignore emails from services at which they don't have an account, and may thus ignore the confirmation email.

\textbf{2. \phaseTwo.} If the victim later decides to create an account with the target service, they might choose to use the federated route to create this account.
Since there is already an account associated with the victim's email address, one plausible design decision could be for the service to merge the two accounts, either intentionally or unintentionally.
At this point the service might notify the victim that there was already an account associated with that email address.

If the victim notices something is amiss, they might thwart the attack by resetting the password, which will succeed because only the victim will receive the password reset email.
However, if the notification is ignored, or worse, not shown, the victim might start using this ``newly-created'' account without resetting the password.
If the service again asks the victim to verify their email address, they might proceed as they would be expecting this.

\textbf{3. \phaseThree.} As shown in state \textbf{S$_{8}$} of Figure~\ref{fig:attacks1}, the victim's IdP is correctly associated with the account, but the password is still that chosen by the attacker.
The attacker can thus directly sign in to the victim's account using this password, whilst the victim continues to sign in via SSO.
In terms of stealthiness, the victim may not notice the attack unless the service notifies users about sign-in events (e.g., sends a notification, or displays the last sign-in time, etc.).

%% file: attack2.tex
\subsection{\unexpSess Attack}
\label{sec:atk2}

This attack exploits a vulnerability in which authenticated users are not signed out of an account when the password is reset.
This allows the attacker to retain access to a pre-hijacked account even after the victim resets the password.

\textbf{Preconditions.} The target service must allow a user to reset the account's password (e.g., by sending a password reset email to the email address associated with the account).
The service must also allow multiple concurrent sessions.

\textbf{1. \phaseOne.} The attacker creates an account at the target service using the victim's email address.
The attacker then signs in to the created account and keeps the session active indefinitely.
Depending on the service, the attacker might be able to automate the task of keeping the session active (e.g., running a script that periodically performs an action).

\textbf{2. \phaseTwo.} Unaware of the attacker's actions, the victim tries to create an account at the target service using their email address.
However, since there is already an account associated with the email address, the service should block the account creation.
The victim may assume that they had previously created an account, but will be unable to sign in given the attacker-chosen password.
The victim will then proceed to reset the password and begin using the account.

\textbf{3. \phaseThree.} If the service does not invalidate all active sessions when the password is reset, it could reach state \textbf{S$_{9}$} of Figure~\ref{fig:attacks1}.
Although the password is only known to the victim, the attacker still has an active session, and can thus continue to access the account as long as that session is maintained.
The victim may not notice the attack unless the service provides some indication of current active sessions.

Another variant of this attack is where the victim instead attempts to create an account via the federated route (as in the \classFedMerge Attack), but the service prevents the account creation on the basis that there is already an account associated with that email address.
The victim may decide to reset the password, and the attack would proceed as above.

%% file: attack3.tex
\subsection{\trojId Attack}
\label{sec:atk3}

This attack combines actions from the \classFedMerge and \unexpSess attacks.
The attacker creates a pre-hijacked account using the victim's email address, but then associates the account with the attacker's IdP account for federated authentication.
When the victim resets the password (as in the \unexpSess Attack), the attacker can still access the account via the federated authentication route.

\textbf{Preconditions.}
The target service must support both classic and federated authentication.
However, unlike in the \classFedMerge Attack, users need not be able to \emph{create} accounts via the federated route -- it is sufficient that one or more federated identities can be \emph{added} to an existing account.
The attacker must have an account with an IdP that is supported by the service, and from which federated identities can be added to existing accounts at the service.
As before, the target service must provide password reset functionality. 

\textbf{1. \phaseOne.} The attacker creates an account at the target service using the victim's email address and a password of the attacker's choosing.
The attacker then associates their own federated identity to the account, such that the attacker can sign in via either SSO or the chosen password.

\textbf{2. \phaseTwo.} When the victim attempts to create an account with the target service, the account creation will fail as there is already an account associated with that email address.
Although unable to sign in (due to the attacker-chosen password), the victim might think they had previously created the account and proceed to reset the password.
This will succeed, and the victim may start using the account as normal.

\textbf{3. \phaseThree.} As shown in state \textbf{S$_{10}$} of Figure~\ref{fig:attacks1}, the attacker's federated identity (denoted as \emph{attackerIdPId}), is still associated with the account, allowing the attacker to sign in to the victim's account via the federated authentication route.
In terms of stealthiness, the victim does not lose access to the account, so may not suspect that there has been an attack.

\textbf{Alternative identifier variant.} A variant of this attack exists where the service 1) allows users to add an alternative identifier, such as a phone number or second email address, and 2) allows users to reset the account password or request a one-time sign-in link via this alternative identifier.
Specifically, during the pre-hijack phase, the attacker associates an attacker-controlled alternative identifier with the account.
The attacker can leverage simple deception techniques to reduce the likelihood of the victim noticing this identifier, such as choosing an email address that is similar to that of the victim (e.g., victim@example.com and victm@example.com), or an address that looks like it is part of the service (e.g., target-service@example.com).
In the attack phase, the attacker regains access to the account by requesting a password reset or one-time sign-in link via this alternative identifier.
Resetting the password would be noticeable, as the victim would lose access to the account, but accessing the account via a one-time sign-in link may allow the attacker to remain undetected.

%% file: attack4.tex
\subsection{\unexpEmail Attack}
\label{sec:atk5}

This attack exploits a potential vulnerability arising from the failure to invalidate email-change capability URLs upon password reset.
The attacker creates an account using the victim's email address and begins the process to change the email address associated with the account to the attacker's own email address, but does not complete this process.
After the victim has recovered the account, the attacker completes the change-of-email process to take control of the account.

\textbf{Preconditions.} The target service must allow users to change the email address associated with an account.
In particular, the service must attempt to verify that the user owns the new email address by sending some type of capability to the new email address (e.g., a link to click or a code to enter).
Furthermore, this sent capability must have a reasonably long validity period (e.g., days).

\textbf{1. \phaseOne.} The attacker creates an account at the target service using the victim's email address.
The attacker then starts the process to change the email address to the attacker's own email address.
The service will send a verification email to the attacker's email address, but the attacker does not yet confirm the change.
Without confirmation, the email change will not proceed, and the account will still be associated with the victim's email, as shown in state \textbf{S$_{6}$} of Figure~\ref{fig:attacks1}.

\textbf{2. \phaseTwo.} Unaware of the attacker's actions, the victim tries to create an account at the target service using their email address.
As before, the account creation will fail because of the existing account associated with the victim's email address.
Similarly, the victim will proceed to reset the password, which is possible since the account is still primarily associated with the victim's email address.
The victim would then proceed to use the account as normal.

\textbf{3. \phaseThree.} At a later point in time, the attacker confirms the pending change-of-email using the capability sent in the verification email from the pre-hijack phase.
If this succeeds, it will change the email address associated with the account, as shown in state \textbf{S$_{11}$} of Figure~\ref{fig:attacks1}.
The attacker can then use the email-based password reset feature to reset the password to an attacker-controlled value.

After the password reset, the victim will lose access to the account, thus lowering the stealthiness of this attack.
However, similarly to the \trojId Attack, if users can request one-time sign-in links sent to their email addresses, the attacker could use this functionality instead of resetting the password.
This allows the victim to continue using the account and thus improves the attack's stealthiness.

\textbf{CSRF variant.} In some services, the attacker might be signed out of the account at the end of the victim action phase (i.e., after the victim resets the password).
Furthermore, these services may require users to be signed in when visiting the email-change capability URL, which would prevent the above attack.
However, in a variant of this attack, the attacker could forge an HTTP request to the email-change capability URL by tricking the victim into visiting this URL while the victim is logged in at the recovered account (e.g., through a CSRF attack or Click-bait).
However, if the victim needs to perform an action upon visiting the capability URL (e.g., entering a confirmation code), this will not be possible.

%% file: attack5.tex
\subsection{\nonverifIdp Attack}
\label{sec:atk6}

This attack is the mirror image of the \classFedMerge Attack, where the attacker created the account using the classic route and the victim used the federated route.
In this attack, the attacker leverages an IdP that does not verify ownership of an email address when creating a federated identity.
We refer to this as a \emph{non-verifying} IdP.
Using this non-verifying IdP, the attacker creates an account with the target service and waits for the victim to create an account using the classic route.
If the service incorrectly combines these two accounts based on the email address, the attacker will be able to access the victim's account.

\textbf{Preconditions.} The target service must support both classic and federated account creation and sign in.
It must also support at least one non-verifying IdP, thus allowing the attacker to create a federated identity based on the victim's email address.
Alternatively, the service must allow users to integrate their own custom IdPs.
This feature is more common for business accounts as it allows users of a particular organization to sign in through their organization's IdP.
The availability of cloud-based IdP services such as OneLogin~\cite{OneLogin} and Okta~\cite{Okta} makes this step easy for the attacker.
We found that these IdPs did not perform email verification for \emph{test accounts}, yet these accounts could still be used as federated identities at other services.

\textbf{1. \phaseOne.} The attacker creates a federated identity (i.e., an IdP account) using the victim's email address at the non-verifying IdP.
The attacker then uses this federated identity and the custom IdP authentication feature to create an account with the victim's email address at the target service.
If the service performed its own verification for email addresses associated with federated identities, it would block this attack.
However, some services simply assume that IdPs have performed email verification, possibly to minimize user friction during account creation.

\textbf{2. \phaseTwo.} The victim subsequently creates an account at the target service via the classic route using their email address and victim-chosen password.
Since an account already exists for that email address, the target service might just combine this with the account created in the pre-hijack phase (similarly to the \classFedMerge Attack).
Thinking that this was a newly-created account, the victim might proceed to use it as normal.

\textbf{3. \phaseThree.} As shown in state \textbf{S$_{12}$} of Figure~\ref{fig:attacks2}, the attacker's federated identity is still associated with the victim's account, allowing the attacker to sign in via the non-verifying IdP.

In this attack, the services incorrectly trust the IdP to perform email-verification.
The fact that the IdP does not verify email addresses could be intentional (e.g., a business decision) or unintentional.
For example, a previously-discovered vulnerability~\cite{PelesLinkedinSpoofedMe} in widely-used IdPs, which has now been fixed, allowed an attacker to create accounts with the victims' email addresses, which could then be used to create accounts with various target services.
Similar techniques could be repurposed to perform account pre-hijacking attacks.

Although we described this attack using a non-verifying IdP, it would also work equally well with a malicious IdP.
Mainka et al.~\cite{Mainka2016MalIdp} investigated the dangers of malicious IdPs, but did not consider using a malicious IdP for account pre-hijacking attacks.

\textbf{Federated-Federated variant.} A variant of this attack is that, in the \emph{victim action} phase, the victim creates their account at the target service using the federated route instead of the classic route.
This only works if the target service allows more than one IdP to be associated with a single account.
As shown in state \textbf{S$_{13}$} of Figure~\ref{fig:attacks2}, this results in both the victim's and attacker's federated identities being associated with the account.
In the \emph{attack} phase, the attacker can therefore sign in via the federated route, and the victim will continue to sign in via their own IdP account.

%% file: attack7.tex
\subsection{Email Verification Trick}
\label{sec:emailVerifVariation}

When mounting some of the above attacks, the target service may require verification of the supplied email address.
For example, the service could send an email to the supplied address containing a unique URL for the user to visit, or a confirmation code which the user must provide to the service.
Since the attacker does not have access to the victim's email account, this could help to mitigate attacks in which the attacker needs to create an account using the victim's email address (i.e., all except the \nonverifIdp Attack).

However, the attacker may be able to circumvent this check by abusing the change-of-email functionality provided by many services.
Specifically, the attacker starts by creating an account at the target service using an identifier they control. For instance, the email address of the attacker (state \textbf{S$_{4}$} of Figure~\ref{fig:attacks2}).
During account creation, the service requests verification of the identifier, which the attacker can perform since they control the identifier.
After the account has been created, the attacker proceeds to change the primary email address to that of the victim (state \textbf{S$_{7}$} of Figure~\ref{fig:attacks2}).
If the service does not perform another email verification at this point (or if it associates the victim's email address to the account before the verification), the attacker is once again in a position to mount any of the attacks described above (e.g., the \classFedMerge Attack, resulting in state \textbf{S$_{14}$} of Figure~\ref{fig:attacks2}).

%% file: experiments.tex
\section{Experiments \& Results}
\label{sec:exp}

To measure the prevalence of vulnerabilities that could lead to the account pre-hijacking attacks described in the previous section, we analyzed some of the most popular websites, based on the Alexa global website rankings~\cite{alexatop500}.
In Section~\ref{sec:methodology}, we discuss our experimental methodology, including the criteria we used to identify specific services of interest, the set of actions we undertook for each service, and the ethical considerations which guided our analysis.
In Section~\ref{sec:results} we present a summary of our results and in Section~\ref{sec:cases} we discuss five case studies  of prominent services that are illustrative of the types of attacks we observed during our analysis.

\subsection{Methodology}
\label{sec:methodology}
Due to the varied nature of the websites, we relied on manual (human) analysis.
Although this limited the number of services we could analyze, it provided the most accurate results because it is similar to how a potential attacker would analyze each service.
We discuss the potential for automating the detection of such vulnerabilities in Section~\ref{sec:autodetect}.

\subsubsection{Selection criteria}
\label{sec:selection}
Given the manual effort required, we limited our analysis to the top \nSites{} websites from the Alexa global website rankings~\cite{alexatop500}.
Of these, \nSitesApplicable{} supported account creation, and out of these, we were able to test \nSitesTested{} for at least one attack.
We could not test the remaining \nSitesSkipped{} websites due to language barriers, specific requirements for account creation (e.g., needing to provide a phone number in a specific country), and other errors.
We also omitted websites where a very similar website had already been tested (e.g., we tested amazon.com but skipped amazon.in as the latter is a localized version of the former).
The summary of reasons for which websites could not be tested is shown in Table~\ref{tab:skipped}.
Although this is by no means a representative sample of all websites, it is reasonable to assume that the most popular websites expend significant effort in preventing user account hijacking, and thus our results are likely a lower bound (i.e., conservative estimate) of the prevalence of such vulnerabilities across all websites.

\input{tables/tbl_skipped}

\subsubsection{Analysis actions}
We specifically focused on email addresses as this is the most common type of unique user identifier. 
As per the preconditions, at the beginning of the attack, the victim's email address should not be associated with any account at the target website.
This made it necessary to create multiple fresh accounts on each website for each test.
In order to identify whether a website met the preconditions of each attack, we created an account and explored and documented the features provided by the service.
Specifically, we looked for the following types of functionality:
\begin{itemize}
    \item Create an account via the federated route;
    \item Use a custom IdP for federated account creation;
    \item Associate a federated identity with an account;
    \item Change the email address associated with an account;
\end{itemize}
During our experiments on the Alexa top \nSites{} websites, we encountered \nIdPs unique IdPs. Out of them, Google was the most popular (integrated in \nSitesWithGoogleIdP websites), followed by Facebook (\nSitesWithFbIdP websites) and Apple (\nSitesWithAppleIdP websites). For our experiments, we preferred the IdPs with the most frequency (i.e., whenever we found websites supporting multiple IdPs, we ran our tests using the most-popular IdP). For the attacks involving email addresses, we preferred Google's mail service, as we could also use the same account for federated authentication.

Whenever a website appeared to be vulnerable, we repeated the test with another fresh set of accounts to avoid false positives and collect additional information for reporting purposes (e.g., creating video-recordings of the procedure).
Additionally, we found that websites use several different channels and templates for reporting security vulnerabilities.
Some websites used third-party vulnerability disclosure services, such as HackerOne~\cite{HackerOne} and Bugcrowd~\cite{BugCrowd}, whilst others provided dedicated vulnerability reporting forms and security email addresses.
We identified these channels and are in the process of reporting our findings to the affected vendors.

\subsubsection{Ethical considerations}
We took multiple precautions to ensure that our experiments were conducted ethically and the results disclosed responsibly.

Firstly, we ensured that the impact of our experiments did not go beyond the users accounts we had created for this purpose.
We limited our experiments to targeted attacks, and thus did not inadvertently disclose sensitive data belonging to other users of the website.

Secondly, we did not leverage any automatic tools that could have sent large numbers of requests, as this might have adversely affected the performance of the website for other users.
The manual nature of our experiments was equivalent to a single legitimate user interacting with the website.
This also meant that we did not trigger any bot detection alerts (e.g.,~\cite{Azad2020Dimva}), which improved the accuracy of our results.

Finally, for each vulnerable website, we submitted a detailed report to the affected vendor (as shown in Table~\ref{tab:resultsfull}). 
In some cases, we have already received acknowledgments and bug bounties for these reports.

In all cases, we ensured that the affected vendors have had at least 90 days to remediate the reported vulnerabilities prior to the publication of this paper.
Additionally, for the case studies presented in Section~\ref{sec:cases}, we obtained permission to publish the details of the vulnerabilities from the affected vendors.

%% file: tables/tbl_skipped.tex
\begin{table}[t]
	\small
	\centering
	\begin{tabularx}{\columnwidth}{X c} 
		\toprule
		\textbf{Reason} & \textbf{Occurrences} \\ 
		\midrule
		Similar website already tested & \nSkippedDupe{}\\
		High-requirements & \nSkippedReq{}\\
		No authentication functionality & \nSkippedNoAuth{}\\
		No email functionality & \nSkippedNoEmail{}\\
		Language barrier & \nSkippedLang{}\\
		Service not reachable & \nSkippedNoReach{}\\
		Other service errors & \nSkippedError{}\\
		\midrule
		\textbf{Total} & \textbf{\nSitesSkipped{}}\\
		\bottomrule
	\end{tabularx}
\caption{Reasons for which services could not be tested.}
\label{tab:skipped}
\end{table}

%% file: results.tex
\subsection{Results \& Observations}
\label{sec:results}

\input{tables/tbl_results}
\input{tables/tbl_results_full}
The summarized results of our analysis are shown in Table~\ref{tab:results} and the detailed list of vulnerable services in Table~\ref{tab:resultsfull}.

In Table~\ref{tab:results}, the second column shows how many services we were able to identify as potentially vulnerable to each type of attack (e.g., they supported account creation, federated identities, etc. as required for each attack). The sum of this column (i.e., \nTests) corresponds to the total number of pre-hijacking attacks we performed.

The third column shows how many services, out of those that were potentially vulnerable, were actually vulnerable to each attack.
As shown in Table~\ref{tab:results}, we identified at least \nVulns{} individual vulnerabilities across all attack types and services.
Note that this does not mean that the remaining services from the top 150 list were invulnerable, as we were not able to test all services due to language barriers or country-specific requirements, as explained in Section~\ref{sec:selection}.

The \unexpSess Attack had the highest number of potentially vulnerable services.
This is to be expected since this attack has minimal requirements (e.g., does not require the service to support federated authentication) and is theoretically applicable to any service that uses the concept of a session.
We also observed that some services that were not vulnerable to the \classFedMerge Attack might still be vulnerable to a variant of the \unexpSess Attack after the former has been blocked.
However, we did not include these in Table~\ref{tab:results} to avoid potential double-counting.

We found  that a similar number of services were vulnerable to the \classFedMerge, \trojId, and \unexpEmail Attacks, when including those for which the Email Verification trick (Section~\ref{sec:emailVerifVariation}) was used.
However, as shown in Table~\ref{tab:resultsfull}, it is not usually the case that the same service is vulnerable to all three of these attacks.

The \nonverifIdp Attack was far less prevalent as it requires the service to support user-specified IdPs for federated authentication.
However, we expect that our results are an under-approximation as some services might provide this functionality as part of a premium plan, which was not always tested.
Nevertheless, out of the three services that we identified as being potentially vulnerable, at least one was vulnerable to this attack.

It is important to note that, at the time of testing, these results were conservative lower bounds, as it is possible that we might have missed some attacks.
On the other hand, we hope that the number of vulnerable services will have significantly decreased due to our responsible disclosures.

%% file: tables/tbl_results.tex
\begin{table}[t]
	\small
	\centering
\begin{tabularx}{\columnwidth}{X x{17mm} x{17mm}}
	\toprule
	\textbf{Attack} & \textbf{Potentially vulnerable} & \textbf{Vulnerable} \\
	\midrule
	\classFedMerge & \nAtkOneTested{} & \nAtkOneVuln{} \\
	\unexpSess & \nAtkTwoTested{} & \nAtkTwoVuln{} \\
	\trojId & \nAtkThrTested{} & \nAtkThrVuln{} \\
	\unexpEmail & \nAtkFouTested{} & \nAtkFouVuln{} \\
	\nonverifIdp & \nAtkFivTested{} & \nAtkFivVuln{} \\
	\midrule
	\textbf{Total} & \textbf{\nTests} & \textbf{\nVulns}\\
	\bottomrule
\end{tabularx}
\caption{Summary of vulnerabilities identified during our testing (January -- June 2021). The \textbf{Potentially vulnerable} column shows how many services we were able to test that could potentially have been vulnerable to each attack (e.g., providing all the necessary functionality), whilst the \textbf{Vulnerable} column shows how many were vulnerable to each attack. Some services were vulnerable to multiple attacks and the detailed results are presented in Table~\ref{tab:resultsfull}.}
\label{tab:results}
\end{table}

%% file: tables/tbl_results_full.tex
\begin{table*}[p]
	\small
	\centering
	\renewcommand{\arraystretch}{0.9}
	\begin{tabularx}{\textwidth}{ X x{15mm} x{15mm} x{15mm} x{15mm} x{15mm} p{25mm}}
		\toprule
		\textbf{Type of Service} & \textbf{\classFedMerge} & \textbf{\unexpSess} & \textbf{\trojId} & \textbf{\unexpEmail} & \textbf{\nonverifIdp} & \textbf{Disclosure Date \& Channel} \\
		\midrule
		video conferencing&	\vuln	&	\tested	&	\tested	&	\tested	&	\vuln	& Mar '21, \HOneP \\ 
		\midrule[0.01mm]
		photo sharing social network&	\tested	&	\tested	&	\vuln	&	\tested	&	\na	&	Jul `21, \SecFo \\ 
		\midrule[0.01mm]
		news and entertainment&	\vuln	&	\tested	&	\tested	&	\tested	&	\na	&	Sept `21, \GEm \\ 
		\midrule[0.01mm]
		e-commerce&	\tested	&	\tested	&	\vuln	&	\tested	&	\na	&	Sept `21, \SecFo\\ 
		\midrule[0.01mm]
		software company&	\tested	&	\vuln	&	\vuln	&	\tested	&	\na	&	Sept `21, \HOne\\ 
		\midrule[0.01mm]
		professional social network&	\tested	&	\vuln	&	\vuln	&	\vuln	&	\na	&	Jun `21, \SecEm\\ 
		\midrule[0.01mm]
		cloud file storage&	\tested	&	\tested	&	\tested	&	\vulnC	&	\na	&	Jun `21, \HOne\\ 
		\midrule[0.01mm]
		job search&	\tested	&	\tested	&	\tested	&	\vuln	&	\na	&	Sept `21, \BC\\ 
		\midrule[0.01mm]
		blog hosting platform&	\tested	&	\vuln	&	\vuln	&	\vulnC	&	\na	&	Jun `21, \HOne\\ 
		\midrule[0.01mm]
		online learning&	\tested	&	\tested	&	\tested	&	\vulnC	&	\na	&	Sept `21, \SecEm\\ 
		\midrule[0.01mm]
		e-commerce&	\vuln	&	\tested	&	\vuln	&	\tested	&	\na	&	Jul `21, \BC\\ 
		\midrule[0.01mm]
		graphics sharing&	\vuln	&	\tested	&	\tested	&	\vuln	&	\na	&	Jul `21, \SecEm\\ 
		\midrule[0.01mm]
		freelancing platform&	\tested	&	\vuln	&	\tested	&	\tested	&	\na	&	Sept `21, \BC\\ 
		\midrule[0.01mm]
		e-commerce&	\na	&	\vuln	&	\na	&	\tested	&	\na	&	Sept `21, \SecEm\\ 
		\midrule[0.01mm]
		music streaming&	\tested	&	\vuln	&	\vuln	&	\tested	&	\na	&	Jun `21, \HOne\\ 
		\midrule[0.01mm]
		cryptocurrency&	\na	&	\vuln	&	\na	&	\na	&	\na	&	Sept `21, \SecFo \\ 
		\midrule[0.01mm]
		sports news&	\na	&	\vuln	&	\na	&	\na	&	\na	&	Sept `21, \SecFo\\ 
		\midrule[0.01mm]
		adult entertainment&	\na	&	\vuln	&	\na	&	\vuln	&	\na	&	Sept `21, \HOne\\ 
		\midrule[0.01mm]
		real estate&	\vuln	&	\tested	&	\tested	&	\tested	&	\na	&	Jun `21, \HOne\\ 
		\midrule[0.01mm]
		collaboration/productivity&	\vuln	&	\vuln	&	\tested	&	\tested	&	\na	&	Jul `21, \Fed\\ 
		\midrule[0.01mm]
		image sharing&	\vuln	&	\vuln	&	\na	&	\tested	&	\na	&	Jul `21, \HOne\\ 
		\midrule[0.01mm]
		online learning/teaching&	\vuln	&	\tested	&	\tested	&	\tested	&	\na	&	Sept `21, \HOne\\ 
		\midrule[0.01mm]
		productivity tool&	\vuln	&	\tested	&	\tested	&	\vuln	&	\na	&	Sept `21, \HOne\\ 
		\midrule[0.01mm]
		news&	\vuln	&	\vuln	&	\vuln	&	\tested	&	\na	&	Jul `21, \HOneP\\ 
		\midrule[0.01mm]
		news and entertainment&	\na	&	\vuln	&	\na	&	\tested	&	\na	& Sept `21, \SecEm\\ 
		\midrule[0.01mm]
		document management&	\tested	&	\vuln	&	\vuln	&	\tested	&	\na	&	Sept `21, \SecFo\\ 
		\midrule[0.01mm]
		microblogging social network&	\na	&	\vulnS	&	\na	&	\tested	&	\na	&	Sept `21, \HOne\\ 
		\midrule[0.01mm]
		video hosting&	\tested	&	\tested	&	\vuln	&	\tested	&	\na	&	Sept `21, \HOne\\ 
		\midrule[0.01mm]
		music sharing&	\vuln	&	\tested	&	\vuln	&	\vuln	&	\na	&	Jun `21, \BC\\ 
		\midrule[0.01mm]
		Internet tools&	\tested	&	\vulnS	&	\tested	&	\tested	&	\na	&	Sept `21, \GEm\\ 
		\midrule[0.01mm]
		travel reservation&	\tested	&	\vulnS	&	\tested	&	\vuln	&	\na	&	Sept `21, \HOneP\\ 
		\midrule[0.01mm]
		adult entertainment&	\vuln	&	\tested	&	\tested	&	\tested	&	\na	&	Sept `21, \GFo\\ 
		\midrule[0.01mm]
		Internet tools&	\na	&	\vulnS	&	\na	&	\vuln	&	\na	&	Sept `21, \Tw\\ 
		\midrule[0.01mm]
		financial services&	\vuln	&	\tested	&	\tested	&	\tested	&	\na	&	Sept `21, \SecFo\\ 
		\midrule[0.01mm]
		online learning&	\tested	&	\vulnS	&	\vuln	&	\tested	&	\na	&	Sept `21, \GFo\\ 
		\midrule
		\textbf{Total vulnerable}	&	\textbf{13}	&	\textbf{19}	&	\textbf{12}	&	\textbf{11}	&	\textbf{1}	&	\\
		\bottomrule
	\end{tabularx}
	\caption{Vulnerabilities identified during our testing (January -- June 2021). 
	For each service, 
	\na means the vulnerability is not applicable, 
	\tested means the service was tested and found \emph{not} to be vulnerable, 
	and \vuln means the service was found to be vulnerable. 
	Additionally,\vulnC means a CSRF attack is necessary in the Attack phase and\vulnS means that the attack could be carried out without resetting the victim's password (e.g., via a one-time sign-in link).
	We disclosed these vulnerabilities to the services via their respective disclosure channels, including \HOneFull (\HOne),
	\SecFoFull (\SecFo), \SecEmFull (\SecEm), \BCFull (\BC), \HOnePFull (\HOneP), \GEmFull (\GEm), \GFoFull (\GFo), \FedFull (\Fed), and \TwFull (\Tw).}
	\label{tab:resultsfull}
	\end{table*}

%% file: case-studies.tex
\subsection{Case Studies}
\label{sec:cases}

In this section we present five case studies that illustrate how pre-hijacking attacks could be carried out against well-known services.
All the vulnerabilities described in this section have been responsibly disclosed to the respective vendors.

\subsubsection{Dropbox} 
\label{sec:dropbox}
We found that the Dropbox website was vulnerable to a variant of the \unexpEmail Attack.

\begin{figure*}
	\begin{subfigure}{.5\textwidth}
		\centering
		\includegraphics[width=.8\linewidth]{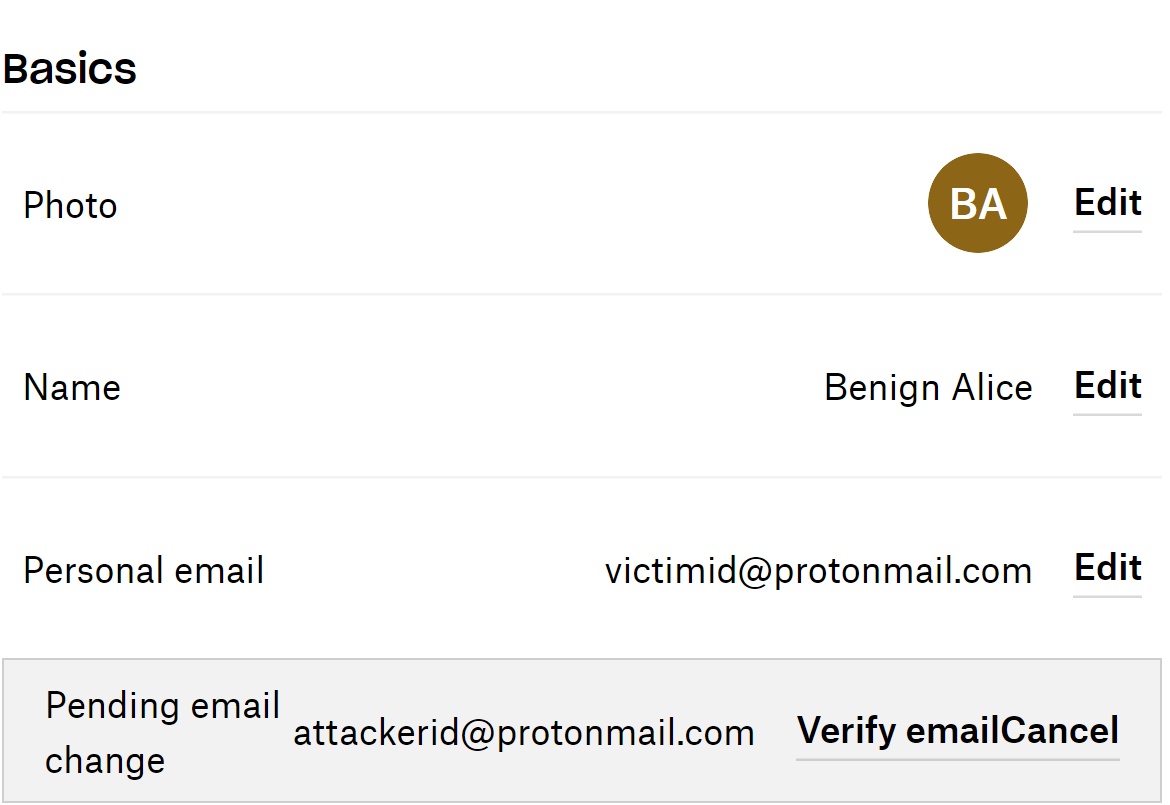}
		\caption{Dropbox UI at the end of \textit{Pre-hijack} phase (Phase 1)}
		\label{fig:dbxphase1}
	\end{subfigure}%
	\begin{subfigure}{.5\textwidth}
		\centering
		\includegraphics[width=.8\linewidth]{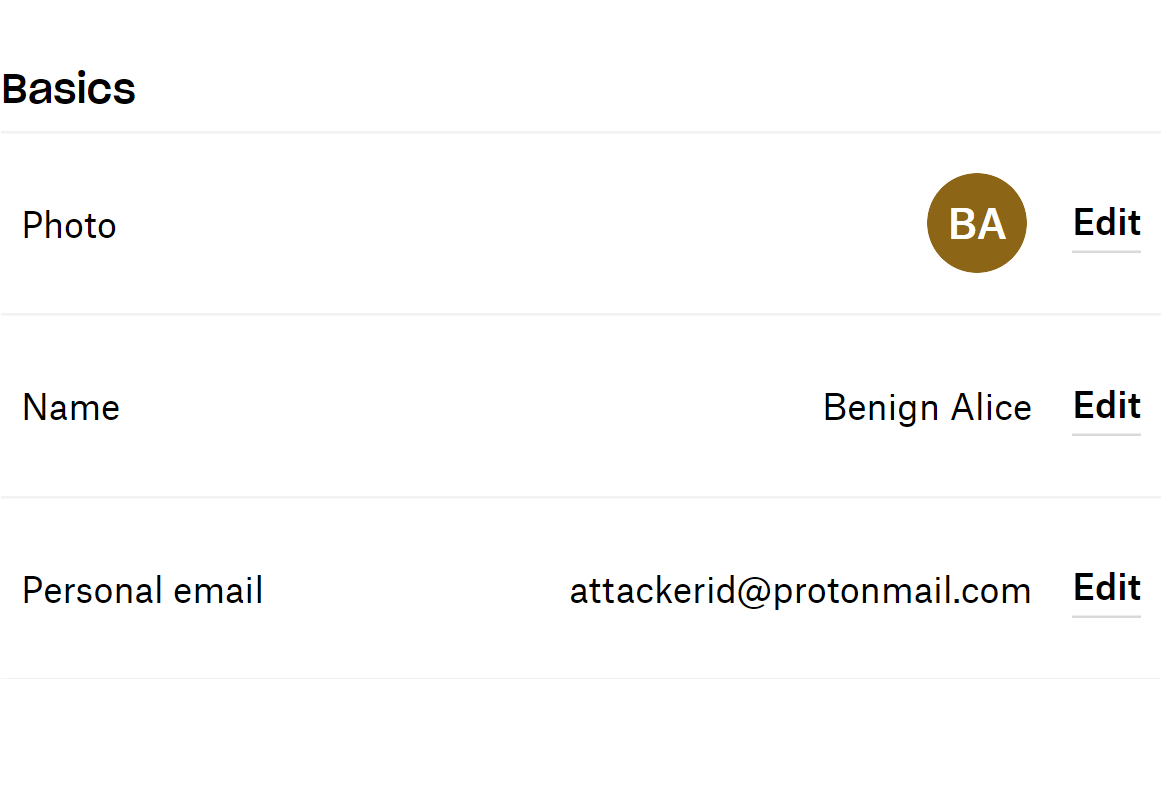}
		\caption{Dropbox UI at the end of \textit{Attack} phase (Phase 3)}
		\label{fig:dbxphase2}
	\end{subfigure}
	\caption{Dropbox UI of the victim's account during the \unexpEmail Attack.}
	\label{fig:dbxatk}
\end{figure*}

\textbf{\unexpEmail Attack.}
As described in Section~\ref{sec:atk5}, the attacker could create an account using the victim's email address.
Dropbox would then send an email to the victim, asking them to confirm their email address.
However, having not signed up for a Dropbox account, the victim might ignore this email because it did not give any instructions as to what they should do if they did not create the account.
The attacker would then start the change-of-email process changing to the attacker's email address, and Dropbox would send a confirmation email to the attacker's email address (transition S$_2$ to S$_6$ in Figure~\ref{fig:attacks1}).
This email contained a URL to confirm the change-of-email, but the attacker would not yet use it.

When the victim tried to create an account using their own email address, this would fail because the email is already associated with an account, and Dropbox would instead ask the victim to sign in to that account.
The victim might then use email-based account recovery and set a new password, causing the attacker to lose access to the account.
As shown in Figure~\ref{fig:dbxphase1}, alert victims might notice the pending email change notification in the user interface (UI) and cancel this.
However, some victims might not notice this.

Some time later, the attacker could make the victim visit the change-of-email confirmation URL (e.g., through a CSRF attack~\cite{CSRF}), which would associate the attacker's email address with the account.
The victim would then see the UI shown in Figure~\ref{fig:dbxphase2}.
The attacker could then use the email-based password reset feature to gain access to the account.

As Dropbox is a cloud-based file storage service and an IdP, a successful attack could allow the attacker to access the victim's private files and sign in to other services where the victim uses Dropbox as an IdP.
However, observant victims might notice that the attacker's email address is also shown in their account (pending confirmation), and might take action to remove this, which would block the attack.
Furthermore, we were not able to test the validity period of the confirmation URL (and the confirmation email did not state a validity period).
The validity period of this URL would also limit the possible window of attack.
We responsibly disclosed our findings to Dropbox via HackerOne in June 2021.

\textbf{Session fixation attack.}
During our experiments, we also discovered a session fixation attack against Dropbox, which allows an attacker to directly sign in to an existing Dropbox account by fixing the session ID~\cite{SessFixation}.
When we reported this issue via HackerOne, it was marked as a duplicate as it had been concurrently reported by another researcher.
Details of the concurrent report are not yet publicly available.

\subsubsection{Instagram}
\label{sec:insta} 
We found that Instagram was vulnerable to the \trojId Attack.

\textbf{\trojId Attack (Alternative identifier variant)}. 
In Instagram, identifier verification is mandatory when creating an account.
Nevertheless, an attacker could create an account using the attacker's phone number, and associate the victim's email address to the created account. 
This would cause a verification email to be sent to the victim's email address.
However, based on our assumptions in Table~\ref{tab:assumptions}, some victims might ignore this email.
When the victim subsequently tried to create an account using their email address, they would find that an account already exists (and might misinterpret this as e.g., being related to the acquisition of Instagram by Facebook, if they already have a Facebook account with the same email address).
The victim might recover the account and start using it.
The attacker would then be able to sign into the account by requesting a one-time sign-in link to be sent to the attacker's phone number.
However, the attack can be thwarted if the victim notices and removes the attacker's phone number from the account.

As Instagram is a social network and an IdP, a successful attacker would be able to access photos and videos shared by the victim and members of their network, and sign in to other services where the victim uses Instagram as an IdP.
The attacker would also be able to read the chats of the victim and impersonate the victim.
When we responsibly disclosed our findings to Instagram in July 2021, they noted that their identifier verification emails include a link to report suspicious sign ups.
However, it is unclear how many victims would take action in this situation, as previous studies (e.g., \cite{Akhawe2013WarningUSENIX}) have shown user-initiated security decisions to be ineffective.
Additionally, Instagram also noted that it is the responsibility of the users to look for Trojan identifiers in their profile.

\subsubsection{LinkedIn}
\label{sec:linkedin}
We found that LinkedIn was potentially vulnerable to the \unexpSess Attack and a variant of the \trojId Attack.

\textbf{\unexpSess Attack.}
This was potentially feasible because LinkedIn did not by default invalidate the active sessions of an account after a password change.
An option for doing this was displayed during the password change procedure, but was not selected by default.
If the victim did not select this option, the account remained vulnerable to this type of attack.
We also noticed that this attack could be performed using the email verification trick (Section~\ref{sec:emailVerifVariation}).

\textbf{\trojId Attack.}
This was potentially feasible because LinkedIn provides the option to associate multiple email addresses with an account.
As described in Section~\ref{sec:atk3}, the attacker creates an account with the victim's email address and then adds their own email address to the account.
This sends an email-change verification URL to the attacker's email address.

After the victim recovers the account and confirms their own email address, any attempt to confirm another email address must be made from an authenticated session.
The attacker thus needs the victim to visit the confirmation URL on the attacker's behalf (e.g., through a CSRF attack).
If successful, the attacker could request a one-time sign in link for this account to be sent to their email address, allowing them to access the account without the victim's password.

As LinkedIn is a professional social network and an IdP, a successful attack could allow the attacker to read the victim's sensitive conversations, impersonate the victim, or sign in as the victim at other services where the victim uses LinkedIn as an IdP.
We reported our findings to LinkedIn in June 2021.
As a result, LinkedIn changed the default behavior to invalidate active sessions after a password change, thus mitigating the \unexpSess Attack.
They also noted that they use multiple defense in depth techniques to minimize the window of vulnerability for \trojId Attacks.
Firstly, the email-change verification URLs are only valid for a limited period of time, forcing the attacker to refresh these regularly.
Secondly, there is only a short time window after the victim's last authentication in which email-change confirmations will be accepted without requiring re-authentication. 
After this window, the victim will be asked to re-authenticate, which would likely raise suspicion.
Finally, LinkedIn uses various anti-abuse controls to prevent the creation of multiple accounts with unconfirmed email addresses.
We discuss these defenses further in Section~\ref{sec:defenseindepth}.

\subsubsection{Wordpress.com}
\label{sec:wordpress}
We found that Wordpress.com was vulnerable to the \unexpSess and \unexpEmail Attacks.

\textbf{\unexpSess Attack.}
In the \emph{\phaseTwo} phase, when the victim tried to create an account with their email address, Wordpress.com notified the victim that an account already exists and provided the option to sign in to the account via a one-time link sent to the victim's email address.
As long as the victim makes use of this option (i.e., does not reset their password), the attacker can maintain their access to the account.
However, even once the victim sets a new password, the attacker's earlier session will not be invalidated, allowing the attacker to retain access potentially indefinitely if the session is kept active.

\textbf{\unexpEmail Attack.}
Similarly to the first case study, in order to successfully execute this attack, the attacker would need to perform a CSRF-like attack in the \emph{\phaseThree} phase.

A successful attack on Wordpress.com would allow the attacker to maliciously modify the websites managed by the victim and sign in to other services where the victim uses Wordpress.com as an IdP. 
When we reported our findings to Wordpress.com via HackerOne in June 2021, the reports were marked as ``Not Applicable''.
These vulnerabilities were not present in the self-hosted version of the Wordpress software because it required all self-registered users to verify their email addresses before allowing them to perform any actions.

\subsubsection{Zoom}
\label{sec:zoom} 
We found that Zoom was vulnerable to the \classFedMerge and \nonverifIdp Attacks.

\begin{figure}
	\centering 
	\includegraphics[width=0.9\columnwidth]{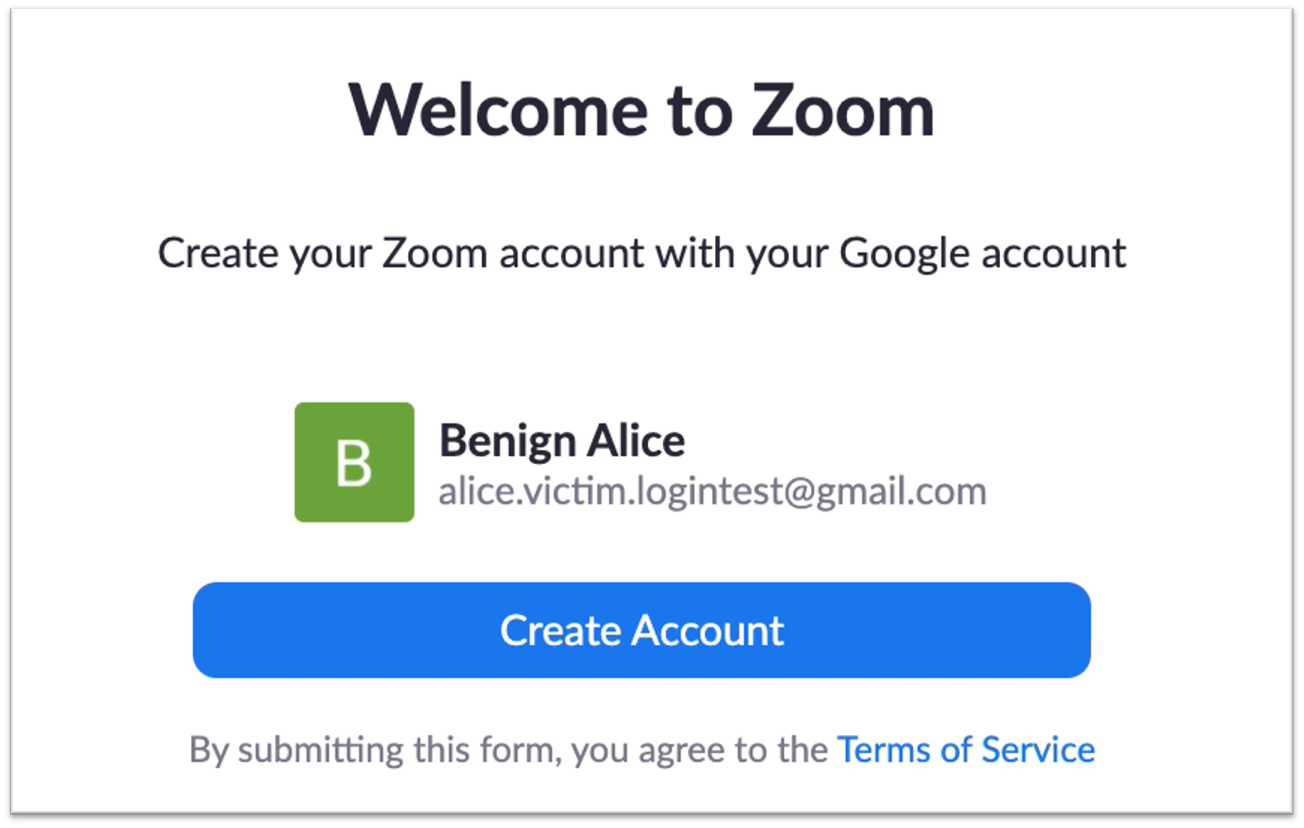}
	\caption{Zoom's UI in the \emph{\phaseTwo} phase of the \classFedMerge Attack. This may lead the victim to believe they are creating a new account instead of being signed in to an attacker-created account.}
	\label{fig:zoomui}
\end{figure}

\textbf{\classFedMerge Attack.}
Although free Zoom accounts require email verification before the account is created, this restriction was not present for paid accounts.
This enables an attacker to abuse the paid account creation process to create an account using the victim's email address and perform the \classFedMerge Attack.
The UI of Zoom when the victim tried to create their account in the \emph{\phaseTwo} phase of this attack is shown in Figure~\ref{fig:zoomui}.
As evident from the figure, the victim would believe they were creating a fresh account, instead of being signed in to the attacker-created account.

\textbf{\nonverifIdp Attack.}
Since Zoom supports custom IdPs, the attacker could use a non-verifying IdP to create a Zoom account with the victim's email address.
For our experiments, we used OneLogin's IdP service~\cite{OneLogin}.
When the victim subsequently came to create a Zoom account with the same email address, Zoom did not notify the victim of the existence of an account with the same email address and instead signed the victim in to the attacker-created account.

Being able to login to the victim's Zoom account would enable the attacker to record the meetings attended by the victim, access the participant details (e.g., attendee names and email addresses) of any meetings hosted by the victim, access the sensitive chat history, impersonate the victim in Zoom chat, and sign in to other services where the victim uses Zoom as an IdP.
When we responsibly disclosed these attacks to Zoom in August 2020 and March 2021, they assessed both reports as \emph{high} severity and fixed the vulnerabilities.

%% file: discussion.tex
\section{Discussion}
\label{sec:discussion}

In this section we discuss account pre-hijacking attacks from the perspective of attackers and defenders, explaining how they might be scaled up to target a larger number of users, as well as identifying the root cause of the vulnerabilities and presenting several defense in depth strategies.
We also sketch out a possible approach for automating the process of scanning for these vulnerabilities, and provide further details about our responsible disclosures to the affected services.

\subsection{Scaling Pre-Hijacking Attacks}
\label{sec:scaling}

In Section~\ref{sec:attacks}, we described the account pre-hijacking attacks from the perspective of an attacker targeting a specific victim on a specific service.
However, it is likely also possible to scale-up the attack to a larger number of users and services.
For example, the attacker could obtain lists of potential identifiers (e.g., email addresses and phone numbers) by scraping social/professional networking services, accessing data from data breach incidents, or by leveraging contact information aggregation services.
The attacker can then try to create accounts for these identifiers at a large number of services, using automated scripts.
Although the account creation may fail on some services, as the victim may already have an account, it is likely to succeed on several other services.
New services that are increasing in popularity are a particularly attractive target for pre-hijacking attacks as victims are less likely to have already created accounts, but are likely to do so in future.
For instance, in 2020, the Zoom video-communications software saw a huge growth in their user base due to the Coronavirus pandemic~\cite{Zoompopularity}.
After creating the accounts, the attacker can leverage automated scripts to periodically check whether the victims have recovered (or started using) the pre-hijacked accounts.
For example, in the \unexpSess{} attack, the attacker can detect account recovery by checking whether the credentials chosen during the pre-hijacking phase can still be used to sign-in to the account.
After the victims start using the pre-hijacked account, the attacker can wait for a suitable period of time (e.g., to allow the victim time to provide sensitive information) before completing the attack.

\subsection{Root Cause \& Mitigation}
\label{sec:mitigations}

\subsubsection{Strict Identifier Verification}
The root cause of all of the attacks identified in the preceding sections is failure to verify ownership of the claimed identifier.
This applies directly to the service itself (as illustrated by the \classFedMerge{}, \unexpSess{}, \trojId{}, and \unexpEmail{} Attacks), as well as to the IdP (see \nonverifIdp{} Attack).
Although many services do perform this type of verification, they often do so asynchronously, allowing the user to use certain features of the account before the identifier has been verified.
Although this might improve usability (reduces user friction during sign up), it leaves the user vulnerable to pre-hijacking attacks.

On the other hand, all of the above attacks could be mitigated if the service or IdP sent a verification email to the user-provided email address and required the verification to be successfully completed \emph{before} allowing any further actions associated with the account.
A similar approach could be used to verify ownership of other types of identifiers, such as using text messages or automated voice calls to confirm ownership of phone numbers.
If the service relies on the IdP to perform verification, it should require a strong guarantee from the IdP that this verification has been performed.
Alternatively, the service could perform its own additional verification, but this adds further friction and negatively effect usability.

\subsubsection{Defense in Depth}
\label{sec:defenseindepth}

In addition to the identifier verification discussed above (or if this is not possible), services can implement the following recommendations to achieve defense in depth.

\textbf{Password resets.}
When the password for an account is reset, the service should perform the following actions:
\begin{itemize}
    \item Sign out all other sessions and invalidate all other authentication tokens for that account. This would mitigate the \unexpSess{} Attack.
    \item Cancel all pending email change actions for that account to mitigate the \unexpEmail{} Attack.
    \item Notify the user of which federated identities, alternate email addresses, and phone numbers are currently linked to the account and ask them to explicitly select which ones to retain (i.e., unlink by default). Alternatively, ask the user to select any identifiers they do not recognize (i.e., retain by default), but this runs the risk of the user ignoring the prompt. Assuming the user acts correctly, this would mitigate the \nonverifIdp{} Attack.
\end{itemize}

\textbf{Merging accounts.}
When a service merges an account created via the classic route with one created via the federated route (or vice-versa), the service must ensure that the user currently controls \emph{both} accounts.
For example, when the user attempts to create an account via the federated route but a classic account already exists for the same email address, the user should be required to provide or reset the password for the classic account. Additionally, the steps we discussed above for strengthening the password-reset process should also be applied.
This would mitigate the \classFedMerge{}, and \nonverifIdp{} Attacks.

\textbf{Email change confirmations.}
When the service sends a capability (e.g., a code or a URL with an embedded authentication token) to confirm a change of email address, the validity period of this capability should be as low as possible, within the constraints of usability, in order to minimize the window of vulnerability for the \unexpEmail{} Attack. However, this will not prevent the attacker from continuously requesting new capabilities whenever the previous ones expire. Therefore, the service must limit the number of times a new capability can be requested from an account to the same, unverified identifier.

\textbf{Unverified-Account Pruning.}
Account pruning refers to the practice of deleting inactive user accounts (e.g., \cite{Pruning}).
Services can apply the same process to unverified user accounts (i.e., accounts pending identifier verification).
Lowering the time threshold for pruning unverified accounts would reduce the window of vulnerability for most pre-hijacking attacks (an exception is the \nonverifIdp Attack).
However, this will not prevent an attacker from creating a new account with the same identifier after the previous account gets pruned. To prevent this, the service should monitor and/or limit the number of times a new account can be created for the same identifier, without the identifier being verified.
However, this in turn could allow the attacker to mount a type of Denial-of-Service (DoS) attack by exhausting the account creation quota of the identifiers of legitimate users.
Therefore, the service can instead reduce the pruning threshold for unverified accounts and leverage bot-detection frameworks to limit the rate at which the attacker can automatically create new accounts.

\textbf{Multi-Factor Authentication (MFA).}
Users can protect themselves from pre-hijacking attacks by activating MFA in their accounts.
Correctly-implemented MFA will prevent the attacker from authenticating to a pre-hijacked account after the victim starts using this account.
In order to prevent the \unexpSess{} Attack, the service must also invalidate any sessions created prior to the activation of MFA.

\textbf{Additional Security Measures.} During our experiments, we observed several additional security measures taken by some of the services we analyzed, and we highlight these here.
Some services included a link in the verification emails sent to the victim to report suspicious activity (e.g., as discussed in the Instagram case study).
Some services notified users of any security-critical changes to their account, such as email/password changes and new devices/location anomalies.
Both of these approaches could help to detect or alert the user to a pre-hijacking attack.
Some services allowed creation of multiple accounts with the same identifier, and thus avoided merging the attacker's and victim's accounts.

\subsection{Automatically Detecting Vulnerabilities}
\label{sec:autodetect}

Although our experiments relied on manual testing, we believe that it should be possible for service owners to automate testing for pre-hijacking vulnerabilities, at least to some extent.
For example, service owners likely have access to a private testing version of the service in which they could \emph{temporarily disable} any anti-automation protection measures.
This should make it possible to write and deploy automated scripts that perform the actions from each phase of the pre-hijacking attacks in Section~\ref{sec:attacks}.
These would likely have to be customized for the specific service (e.g. to automate service-specific processes such as account creation, email change, etc.).
The script could vary the sequence in which these actions are performed to test different patterns of interaction.
This could be used in conjunction with a test oracle that checks the server-side state of the service to identify potential vulnerabilities.
Finally, after executing each test, the state of the test service could be reset to a default value, to accelerate the testing process.
In future, we hope that web application vulnerability scanners might also support this type of semi-automated scanning.

\subsection{Responsible Disclosure}
\label{sec:disclosure}

In Section~\ref{sec:cases}, we showed that account pre-hijacking attacks can have serious consequences including disclosure of sensitive information (e.g., Sections \ref{sec:dropbox} and \ref{sec:linkedin}), website defacement (e.g., Section \ref{sec:wordpress}), and spying on web users (e.g., Section \ref{sec:zoom}).
In light of this, we responsibly disclosed all the \nVulns{} vulnerabilities we identified on \nSitesVuln{} services.
We reported \nVulnsDiscThruPlatfs of the vulnerabilities via third-party vulnerability coordination platforms, including HackerOne~\cite{HackerOne}, Bugcrowd~\cite{BugCrowd}, and Federacy~\cite{Federacy}.
We reached out to a further \nVulnsDiscThruSecFEm companies through their dedicated email addresses and forms for reporting security vulnerabilities.
For those services that lacked dedicated security channels (\nVulnsDiscThruGenFEm), were reached out through their general support emails and forms.
For one service, we could not find any of the above contact channels, so we reached them through their parent company's Twitter profile.
The date on which we disclosed the vulnerabilities to each service is shown in the last column of Table~\ref{tab:resultsfull}.

%% file: related.tex
\section{Related Work}
\label{sec:related}

Many of the techniques we use in our pre-hijacking attacks are inspired by prior work on related classes of attacks.

\textbf{Preemptive account hijacking.}
To the best of our knowledge, Ghasemisharif et al.~\cite{GhasemisharifSSignOff2018USENIX} present the first example of \emph{preemptive account hijacking} in the scientific literature (see Section 5 of \cite{GhasemisharifSSignOff2018USENIX}).
Our work builds upon their example, but assumes a significantly weaker attacker.
Specifically, they consider an attacker who has gained control of the victim's federated identity (e.g., the victim's IdP account).
Although this is certainly possible, as they demonstrate in the same work~\cite{GhasemisharifSSignOff2018USENIX}, it goes beyond the standard web attacker threat model (e.g., \cite{AkhaweFormal2010CSF}). 
Additionally, multi-factor authentication (MFA) at the IdP, and browser-level protection mechanisms such as HTTP Strict-Transport-Security (HSTS)~\cite{Hsts}, make it more difficult for an attacker to execute their attack. 
Nevertheless, their attack is more powerful than any of ours, since once a victim's IdP account has been compromised, this can be used to hijack even the existing RP accounts of the victim.
In concurrent work with ours, Ghasemisharif et al.~\cite{Ghasemisharif2022} also present a tool for automated auditing of session management flaws in SSO, which again assumes that the attacker has compromised the victim's IdP account at the session level.
In contrast, we show that there exists a class of account pre-hijacking attacks that are possible \emph{without compromising the victim's federated identity}.
These attacks are thus significantly easier to execute, as demonstrated by our analysis in Section~\ref{sec:results}.

As we explained in Section \ref{sec:disclosure}, some vendors marked our reports on the \classFedMerge{} Attack as duplicate (indicating that the attack had been reported previously). Upon further investigation, we identified two recent HackerOne reports \cite{HalabiPrehijack2020, DhakalPrehijack2021} that mention a variant of our attack (referred to as \textit{pre-account takeover}). In this variant, it is assumed that the victim will directly try to do federated authentication after the attacker has pre-hijacked the account. In our version of the \classFedMerge{} Attack, we make the assumption that the victim will first try to create an account through federated means before trying to sign in. Nevertheless, these reports indicate that the security bug bounty research community is already-aware of account pre-hijacking attacks.

\textbf{Account hijack with SSO.}
Mainka et al.~\cite{Mainka2016MalIdp} explain how a web attacker could hijack the victim's RP account through a malicious IdP.
Similarly, Peles~\cite{PelesLinkedinSpoofedMe} explores the possibility of an attacker signing in to the existing RP accounts of the victim by leveraging a vulnerable IdP that allows SSO logins before the completion of email verification.
These are closely related to and influenced the techniques we use in our \nonverifIdp{} Attack.
Homakov et al.~\cite{HomakovOAuthCSRF} and Sclafani et al.~\cite{SclafaniOAuthCSRF} discuss attacks that enable an attacker to authenticate to the victim's RP account by connecting the attacker's IdP account to it.
This contributed to the technique we use in our \trojId{} Attack.
Although there is a large body of literature discussing logical vulnerabilities in the login process e.g., \cite{SudhodananMPWA2016NDSS, BaiAuthScan2013NDSS, WangWebSSOBrm2012IEEESP, SunOAuthSec2012CCS, ArmandoSAML2008FMSE}, there has been relatively little focus on the security of account creation, which as we have shown in this work, is also critical for protecting users.

\textbf{Account hijack without SSO.}
Zeller et al.~\cite{ZellerCrossSiteRF2008} present an attack in which the attacker associates their own email address to the victim's account and takes control of the account by requesting a password reset link to be sent to the attacker's email address. Similarly, Innocenti et al.~\cite{Innocenti2021Dimva} discusses attacks based on the password-reset URLs and Lee et al.~\cite{LeePhNum2021} presents the scenario where an attacker owns the recycled phone number of the victim to hijack the accounts of the victim through SMS-based password reset. Although the threat model considered in these works are different from ours, they inspired our \unexpEmail{} and \trojId{} Attacks.
Additionally, the classic account hijack attack using CSRF~\cite{CSRF} talks about the capability of the attacker to sign in to the victim's account by setting an attacker-chosen password on the account.
This attack motivated our discovery of the \classFedMerge Attack.

\textbf{Login CSRF.}
The common element of account pre-hijacking and login CSRF attacks (e.g., \cite{BarthCSRF2008CCS, SudhodananAuthCSRF2017EuroSnP}) is that, in both cases, the attacker attempts to trick the victim to use an attacker-controlled account.
Both classes of attacks thus allow the attacker to spy on the actions performed by the victim while they are signed in to the attacker-controlled account.
However, the main difference is in the way the victim ends up signed in to the attacker-controlled account. 
In login CSRF attacks, the victim is forcefully authenticated to the attacker's account, which may be noticed by observant users.
In contrast, in account pre-hijacking, the victim willingly signs in to what they believe to be their own account, unaware that it has been pre-hijacked.

%% file: conclusion.tex
\section{Conclusion \& Future Work}
\label{sec:concl}
In this paper, we show that there exists an entire class of account pre-hijacking attacks, which ultimately allow an attacker to take control of a victim's account.
We describe \nNovelAtkTypes{} specific attacks.
To measure the prevalence of these vulnerabilities in the wild, we analyzed the top \nSitesTested{} popular services based on the Alexa global rankings.
We found that at least \nSitesVuln{} of these were vulnerable to one or more pre-hijacking attacks, including major services such as Dropbox, Instagram, LinkedIn, Wordpress.com, and Zoom.
We disclosed these vulnerabilities to the affected organizations, some of which have acknowledged the vulnerabilities as being of high severity.
Finally, we analyzed the root cause of these vulnerabilities, and presented a set of security requirements that would mitigate all the vulnerabilities we identified. 

There are several potential avenues for future research related to account pre-hijacking attacks.
Firstly, account pre-hijacking attacks are largely possible because of services not performing strict identifier verification (e.g., to reduce user friction during account creation). 
An empirical evaluation of this topic could explore the precise usability benefits and contrast these against the security risks.
Secondly, automated detection of account pre-hijacking vulnerabilities is a promising avenue for future work, as such tools could extend the evaluation to more websites, mobile applications, and desktop applications. 
Although we briefly sketch a design for this in Section \ref{sec:autodetect}, this could be explored in greater detail. 
Finally, pre-hijacking attacks such as the \unexpEmail Attack show the importance of capability URLs.
Further research is needed on the types of capability URLs in use, their typical validity periods, and the ways in which they might be abused (e.g., \cite{Innocenti2021Dimva}).

%% file: ack.tex
\section{Acknowledgements}
\label{sec:ack}
We thank our shepherd, Emily Stark, and the anonymous reviewers for their valuable feedback. We also thank the vendors who collaborated with us during the responsible disclosure process. This work was supported by a research grant from the Microsoft Security Response Center (MSRC).